\documentclass{aastex}
\usepackage{amsmath}
\usepackage{amssymb}
\usepackage[numberedappendix]{emulateapj5}
\usepackage{epsfig}

\makeatletter

\newenvironment{inlinefigure}{%
\def\@captype{figure}%
\noindent\begin{minipage}{0.999\linewidth}\begin{center}}
{\end{center}\end{minipage}\smallskip}
\makeatother

\newcommand{\kms}{\,{\rm km\,s^{-1}}}
\newcommand{\au}{\,{\rm AU}}
\renewcommand{\day}{\,{\rm d}}
\newcommand{\yr}{\,{\rm yr}}

\newcommand{\pc}{\,{\rm pc}}
\newcommand{\kpc}{\,{\rm kpc}}	
\newcommand{\kev}{\,{\rm keV}}

\newcommand{\ergs}{\,{\rm ergs\,s^{-1}}}

\newcommand{\msun}{\,M_\odot}
\newcommand{\rsun}{\,R_\odot}
\newcommand{\lx}{L_{\rm X}}

\newcommand{\mdot}{\,M_\odot\,{\rm yr}^{-1}}

\newcommand{\sx}{S_{\rm X}}
\newcommand{\flux}{\,{\rm ergs\,s^{-1}\,cm^{-2}}}

\newcommand{\be}{\begin{equation}}
\newcommand{\ee}{\end{equation}}

\newcommand{\sqd}{\,{\rm deg}^2}

\begin{document}

\shorttitle{WIND-ACCRETING NEUTRON STARS}                                    
\shortauthors{PFAHL, RAPPAPORT, PODSIADLOWSKI}


\submitted{Submitted to ApJ}

\title{ON THE POPULATION OF WIND-ACCRETING NEUTRON STARS IN THE GALAXY}

\author{Eric Pfahl\altaffilmark{1}, Saul Rappaport\altaffilmark{1}, 
and Philipp Podsiadlowski\altaffilmark{2}}

\altaffiltext{1}{Center for Space Research, Massachusetts Institute of
Technology, Cambridge, MA, 02139; pfahl@space.mit.edu, sar@mit.edu}
\altaffiltext{2}{Department of Astrophysics, Oxford University, Oxford, OX1 3RH,
England, UK; podsi@astro.ox.ac.uk}


\begin{abstract}

We explore the possibility that neutron stars accreting from the winds of
main-sequence stellar companions account for a significant fraction of
low-luminosity, hard X-ray sources ($\lx \la 10^{35}\ergs$; 1--10\,keV) in the
Galaxy.  This work was motivated by recent {\em Chandra} observations of the
Galactic center by Wang et al.~(2002).  Our calculations indicate that many of
the discrete X-ray sources detected in this survey may be wind-accreting neutron
stars, and that many more may be discovered with deeper X-ray observations.  We
propose that an infrared observing campaign be undertaken to search for the
stellar counterparts of these X-ray sources.

\end{abstract}




\section{INTRODUCTION}\label{sec:intro}

Of all the binary X-ray sources in the Galaxy that contain a neutron star (NS),
the most abundant, though typically not the most luminous, are systems in which
the NS accretes from the wind of a main-sequence stellar companion.  Immediately
following the supernova (SN) explosion that accompanies the formation of the NS,
the companion is, in most cases, relatively unevolved.  Stellar masses ranging
from a few to several tens of solar masses are likely, with corresponding
lifetimes of $\sim 10^7$--$10^8\yr$.  The long phase of wind accretion onto the
NS is a fairly quiet prelude to the more dramatic Roche-lobe overflow that
follows, where the NS may appear as a low/intermediate-mass X-ray binary
(L/IMXB), or, more commonly, as a short-lived high-mass X-ray binary (HMXB)
before being engulfed by its companion.

Well known examples of wind-accreting NSs (WNSs) are present in HMXBs, which are
typically assumed to contain $\ga$ 10-$M_\odot$ stars.  It has long been
suggested \citep[e.g.,][]{rappaport82} that there is a large unobserved Galactic
population of HMXBs with steady X-ray luminosities of $\lx \la 10^{35}\ergs$.
Most importantly for the present work, WNSs in HMXBs have been proposed
\citep[e.g.,][]{ogelman74,koyama89} to account for much of the so-called
``Galactic ridge'' of previously unresolved, hard ($\sim$ 1--10\,keV) X-ray
emission.

The nature and contributions of diffuse and discrete X-ray sources to the X-ray
spectrum and total X-ray luminosity of the Galactic ridge has been a point of
debate for decades.  Prior to {\em Chandra}, X-ray satellites that have surveyed
the Galactic plane (e.g., {\em ASCA}, {\em RXTE}) lacked the angular resolution
and/or sensitivity to identify point sources near the Galactic center.  Wang,
Gotthelf, \& Lang (2002; hereafter, WGL02) recently reported on a
high-resolution survey of a $0.8\degr \times 2\degr$ field about Sgr A$^*$,
carried out with {\em Chandra}/ACIS-I in the energy range of 1--8\,keV.
Throughout most of the observed field, the sensitivity for detecting discrete
sources is $S_{\rm X,min} \sim 10^{-13}\flux$, corresponding to $\lx \sim
10^{33}\ergs$ at the distance ($\sim 8.5\kpc$) of the Galactic center.

WGL02 estimate $\ga$ 500 previously undetected point sources in the {\em
Chandra} survey, the nature of which are unknown.  Sources with X-ray emission
in the 1--3\,keV band are probably within several kiloparsecs of the Sun.  Many
of the sources with energies $\ga 3\kev$, for which the softer X-rays have been
absorbed by the interstellar medium, are nearer to or beyond the Galactic
center.  WGL02 suggest that as many as $\sim 50\%$ of the hard point sources may
be background active galactic nuclei (AGN), although the true fraction may be
closer to $\sim 10$--$20\%$, as estimated from the Chandra Deep Field data
\citep{brandt01}.  This leaves $\ga 100$ Galactic objects that are most likely a
population of binary X-ray sources, including black-hole binaries (BHBs), LMXBs,
cataclysmic variables (CVs), and WNSs.  In the next section, we devote a short
discussion to each candidate class of X-ray binary, and we support the case that
most of the hard ($>$ 3\,keV) Galactic point sources detected by {\em Chandra}
are WNSs.

As we argue below, the field surveyed by {\em Chandra} encompasses roughly 1\%
of the stars in the Galactic disk.  This nontrivial fraction may translate to
$\ga 100$ WNSs (see \S~\ref{sec:bps}), and thus the survey may provide an
important partial census of these objects.  However, even if future observations
reveal that few of the hard sources detected by {\em Chandra} are WNSs, the main
ideas of this paper provide  a foundation for constraining their numbers and
properties.


\section{THE NATURE OF THE POINT SOURCES}\label{sec:nat}

The total numbers for each class of X-ray binary in the solid angle of $1.6\sqd$
observed by {\em Chandra} may be estimated as follows.  We suppose that the
space density of stars in the Galactic disk is given by $n(R,z) \propto
\exp(-R/R_0) \exp(-|z|/z_0)$, where $R$ is the galactocentric radius, and $z$ is
the displacement from the Galactic midplane.  The scale radius, $R_0$, is taken
to be 4\,kpc \citep{kruit87}, and the vertical scale height is $z_0 \ga 100\pc$.
Upon integrating $n(R,z)$ over $1.6\sqd$ through the Galactic center, we find
that the field contains $\la 1\%$ of the Galactic population of hypothetical
X-ray sources.  Thus, the detection by {\em Chandra} of $\ga 100$ Galactic
sources implies a total number of $\ga 10^4$ in the entire Galaxy.  We now
consider each of the candidate X-ray binaries in turn.

Various semi-empirical and theoretical estimates give total numbers of LMXBs and
BHBs (quiescent and active) in the Galaxy at $\sim 10^3$ each
\citep[e.g.,][]{verbunt95,romani98}.  Therefore, it seems that at most of order
$10$ LMXBs and BHBs could have been detected in the WGL02 survey.  This estimate
is consistent with the $\la 20$ LMXBs and one BHB observed in the surveyed
field \citep[see][]{liu01}.

There are probably $\sim 10^6$ CVs in the Galaxy \citep[e.g.,][]{howell01}, and
therefore $\sim 10^4$ CVs in the field surveyed by WGL02.  Most observed CVs
have $\lx \la 10^{32}\ergs$ in the bandpass of 0.1--2.5\,keV, and decreasing
power toward higher energies.  Luminous CVs, with $\lx \sim
10^{33}$--$10^{34}\ergs$, which should comprise only $\sim 1\%$ of the intrinsic
population \citep[e.g.,][]{howell01}, likewise have relatively soft X-ray
spectra.  Therefore, we expect that few CVs are detectable beyond several
kiloparsecs for $S_{\rm X,min} = 10^{-13}\flux$ (1--8\,keV), because of their
low total $\lx$, as well as the heavy interstellar absorption for energies $\la
3\kev$.  Integrating $n(R,z)$ over the $1.6\sqd$ surveyed by {\em Chandra}, to a
distance of $D=3\kpc$ from the Sun, we calculate that several tens of CVs, and
perhaps of order 100, may have been detected.  Therefore, CVs may make up a
sizable fraction of the 1--3-keV sources in the WGL02 mosaic image.

Arguably, the most compelling hypothesis for the hard ($>$ 3\,keV) Galactic
X-ray sources detected by {\em Chandra} is that they are WNSs.  Observed HMXBs
exhibit a wide range of X-ray luminosities ($\sim 10^{33}$-- $10^{38}\ergs$) and
hard, nonthermal X-ray spectra that extend beyond 10\,keV \citep{nagase89}.
Many ($\sim 40\%$) of these systems are transient, with short (less than several
months) outburst phases where $\lx \ga 10^{36}\ergs$, and recurrence times that
are often years or decades.  During its essentially quiescent state, a transient
HMXB may have $\lx \la 10^{35}$, due to steady accretion from the stellar wind
by the NS, such as in the case of the Be/X-ray binary X Per/4U 0352+30
\citep{delgado01}.  Theoretical calculations \citep[e.g.,][]{meurs89} indicate
that tens of thousands of such low-luminosity WNSs may currently inhabit the
Galaxy, implying that hundreds of WNSs may populate the $0.8\degr \times 2\degr$
field about the Galactic center.  In the next section, we use binary population
synthesis to estimate the number of potentially observable WNSs.


\section{THE X-RAY FLUX DISTRIBUTION}\label{sec:bps}

An estimate of the number of WNSs with $\sx > S_{\rm X,min}$ in the field
surveyed by {\em Chandra} requires that we calculate their X-ray flux
distribution.  The X-ray luminosity of an individual source depends upon the
parameters that characterize both the stellar wind and the binary system that
contains the WNS.  We have conducted a Monte Carlo population synthesis study of
the formation of binaries that consist of a NS and a stellar companion.  The
three main steps of our population synthesis calculation are enumerated below.
More detailed descriptions of massive binary stellar evolution and the elements
of our population synthesis code are given in Pfahl, Rappaport, \& Podsiadlowski
(2001; hereafter, PRP01) and \citet{pfahl01c}.

\smallskip

\noindent
1) In the Galactic disk, each binary containing a NS descends from a massive
{\em primordial binary}, where the initially more massive and less massive stars
are hereafter referred to as the {\em primary} and {\em secondary},
respectively.  We take the Galactic formation rate of massive binaries to be
comparable to the core-collapse SN rate, $\mathcal{R}_{\rm SN} \sim
10^{-2}\yr^{-1}$ \citep{cappellaro99}.  The initial primary mass, taken to be
$M_{1i} \ge 8\msun$, is chosen from a power-law IMF, $p(M_{1i}) \propto
M_{1i}^{-2.5}$.  The initial secondary mass, $M_{2i}$, is chosen from a flat
distribution of mass ratios, $q_i\equiv M_{2i}/M_{1i} < 1$.  For simplicity, we
assume that the primordial binaries have circular orbits (see PRP01), and we
choose the initial orbital separation, $a_i$, from a distribution that is flat
in $\log a_i$.

\smallskip

\noindent
2) If the orbit is sufficiently compact ($a_i \la 5-10\au$) that the primary
evolves to fill its Roche lobe, we use analytic formulae (PRP01) to compute the
orbital separation following the subsequent phase of mass transfer.  The mass
ratio and the evolutionary state of the primary at the onset of Roche-lobe
overflow are used to determine whether the mass transfer is {\em stable} or {\em
dynamically unstable}.  Given some critical mass ratio, which we take to be $q_c
=0.5$, the mass transfer is assumed to be stable if $q_i > q_c$ {\em and} the
envelope of the primary is mostly radiative when mass transfer begins, and
dynamically unstable if $q_i < q_c$ {\em or} the primary has a convective
envelope.  We assume that the entire hydrogen-rich envelope of the primary is
removed during mass transfer, whether stable or dynamically unstable, leaving
only the primary's hydrogen-depleted core.  For stable mass transfer, we suppose
that the secondary accretes a fraction, $\beta = 0.75$, of material donated by
the primary, and that the remaining mass escapes the system with a specific
angular momentum that is $\alpha = 1.5$ times the orbital angular momentum per
unit reduced mass.  The orbital separation increases or decreases during stable
mass transfer by a modest factor of $\la 5$--10 for reasonable values of
$\alpha$ and $\beta$ (see PRP01).  We assume in all cases that the secondary is
``rejuvenated'' due to the accretion, so that it emerges following mass transfer
on the ZAMS appropriate for its new mass.  The minimum secondary mass following
{\em stable} mass transfer is $\sim q_c\,8\msun + \beta\,6\msun = 8.5\msun$,
where we have used our reference values of $q_c$ and $\beta$, and $8\msun$ and
$6\msun$ are, respectively, our chosen minimum value of $M_{1i}$ and the
corresponding envelope mass.  Dynamically unstable mass transfer is accompanied
by a common-envelope phase, wherein the secondary experiences a drag that causes
the binary orbit to shrink in a time of $\la 10^3\yr$.  A fraction, $\eta_{\rm
CE} \la 1$, of the initial orbital energy is available to unbind the common
envelope from the system.  If insufficient energy is available, the two stars
will merge.  Otherwise, the envelope of the primary is dispersed, and the
secondary emerges near the ZAMS, without having accreted any mass.  The orbital
separation may be $\sim$ 100 times smaller for binaries that survive dynamically
unstable mass transfer.  A merger results in nearly every case where $q_i < q_c$
and the primary's envelope is radiative when mass transfer starts.  This
implies that almost all systems that survive dynamically unstable mass transfer
must have had initial orbital separations wide enough for the primary to grow to
become a convective red supergiant.

\smallskip

\noindent
3) Upon exhausting its remaining nuclear fuel, the exposed core of the primary
explodes as a Type Ib or Ic SN.  The impulsive mass loss and possibly large
``kick'' to the NS strongly perturb, and may unbind, the binary.  We assume that
both the mass loss and the kick are instantaneous, and that the orientations of
the kicks are distributed isotropically.  Two scenarios are considered for the
distribution of kick speeds.  Observations of {\em isolated} radio pulsars
indicate that the mean NS kick speed may be $\ga 100$--$300\kms$
\citep[e.g.,][]{hansen97,arzoumanian01}.  In our first kick scenario (hereafter,
K1), we apply a Maxwellian distribution in kick speeds with a mean of $\sim
300\kms$ to all NSs.  Our second scenario (K2) was developed \citep{pfahl01c} to
account for a new class of HMXBs with long orbital periods $(P_{\rm orb} \ga
30\day$) and low eccentricities ($e \la 0.2$).  For NS progenitors that are able
to evolve into red supergiants (i.e., single stars and those in wide binaries),
the kick speeds are drawn from a Maxwellian with a mean of $\sim 300\kms$, as in
K1.  Therefore, our arguments in the last paragraph imply that for post-SN
binaries where $M_2 \la8\msun$, the NS has received the conventional large kick.
If the NS progenitor is in a binary system, and its envelope is removed while it
is mostly radiative, which includes all systems where the mass transfer is
stable, we utilize a Maxwellian distribution with a much lower mean of $\sim
30\kms$.  Kick scenario K2 yields a much larger number of post-SN binaries
containing massive companion stars ($\ga 8\msun$), as compared to the number of
surviving systems with $M_2 \la 8\msun$.

\smallskip 

The relevant output of the population synthesis code is the post-SN secondary
mass, $M_2$, semimajor axis, $a$, and eccentricity, $e$, for each binary.  We
neglect tidal circularization during the main-sequence evolution of the
secondary, although this is important for systems with $a(1-e) \la
30$--$40\rsun$.  Our results will not be greatly modified if we include tidal
circularization.

Each WNS accretes from the wind of a relatively unevolved stellar companion.
Winds from early-type stars with masses of $\sim$ 3--$20\msun$ are characterized
by high speeds, $v_w \sim 1000\kms$, and low to moderate mass-loss rates,
$\dot{M}_w \sim 10^{-11}$--10$^{-7}\mdot$.  \citet{kudritzki00} give the
asymptotic wind speeds for stars near the main sequence, which vary from $\sim$
1.5 times the surface escape speed, $v_e$, for stellar masses of $\sim$
3--10$\msun$, and $\sim 2.5 v_e$ for hotter, more massive stars.  The wind
speeds are quite uncertain, however, and we thus consider two simple cases in
our simulations, namely $v_w = v_e$ and $v_w = 2v_e$, independent of $M_2$.  For
$\dot{M}_w$, we utilize the fitting formula of \citet{dejager90}: $\dot{M}_w
\propto M_2^{0.16}L_2^{1.24}R_2^{0.81}$, where $L_2$ and $R_2$ are,
respectively, the luminosity and radius of the secondary.  For both $\dot{M}_w$
and $v_e$, we substitute the time-averaged values of $L_2$ and $R_2$ during the
main-sequence phase --- roughly twice their ZAMS values for a wide range in
$M_2$.

We apply the standard Bondi-Hoyle-Lyttleton accretion scenario
\citep{hoyle41,bondi44} to obtain the X-ray luminosity of each WNS, specialized
to the case where $v_w$ is much larger than the orbital speed of the NS.  We
further assume that the wind is steady and spherically symmetric, so that the
density varies as $\dot{M}_w r^{-2}v_w^{-1}$, where $r$ is the instantaneous
orbital separation.  The orbital time-averaged X-ray luminosity then scales as
$\langle \lx \rangle \propto \epsilon \, \dot{M}_w \, a^{-2} \, v_w^{-4}
\,(1-e^2)^{-1/2}$.  Here, $\epsilon \la 1$ is the efficiency for converting
gravitational energy into hard (1--10\,keV) X-radiation.  For $e = 0$, $a =
0.5\au$, $v_w = 1000\kms$, and $\dot{M}_w = 10^{-8}\mdot$, we find, after
including the multiplicative factors appropriate for an accreting NS, that
$\langle \lx \rangle \sim \epsilon \cdot 10^{33}\ergs$.  In the present work, we
do not take into account the spin history of the NS and the centrifugal
inhibition of accretion \citep[e.g.,][]{stella86}, although this should be
incorporated into a more detailed study.

The X-ray flux distribution of the WNSs is obtained by convolving the distance
and X-ray luminosity distributions, which we assume are independent in this
study. Since the probability of observing a certain WNS should be roughly
proportional to the main-sequence lifetime, $\tau_{\rm MS}(M_2)$, of the
secondary, we compute the distribution of intrinsic X-ray luminosities by
accumulating the values of $\tau_{\rm MS}(M_2)$ for each bin in $\langle \lx
\rangle$.  Using the same exponential form given in \S~\ref{sec:nat} for the
space density, $n(R,z)$, of WNSs, we obtain the distribution of distances, $D$,
from the Sun, for sources located within a solid angle of $1.6\sqd$ centered on
the Galactic center.  We adopt fixed values of $R_0 = 4\kpc$ and $z_0 = 200\pc$
as being typical for the post-SN binaries.  In reality, the effects of SN mass
loss and NS kicks are such that, for a given kick speed, binaries with
lower-mass secondaries will receive larger systemic velocities and reach greater
distances from the Galactic midplane.  A more detailed investigation should
include the dynamical evolution of post-SN binaries in the Galaxy.

Our binary population synthesis calculation yields the formation efficiency,
$\mathcal{F}_{\rm FE}$, for post-SN binaries with $M_2 > 3\msun$ (a somewhat
arbitrary minimum mass), where the remainder of the massive binaries have
ultimately merged or have been disrupted due to the SN.  The secondary stars in
the surviving systems have a mean main-sequence lifetime of $\langle \tau_{\rm
MS} \rangle$, so that the total number of WNSs in the Galaxy is $\sim \langle
\tau_{\rm MS} \rangle \mathcal{R}_{\rm SN} \mathcal{F}_{\rm FE}$.  Multiplying
this total Galactic number by the fraction, $\mathcal{F}_\Omega$, of systems
encompassed by the observed solid angle, we obtain the intrinsic number of WNS
in the surveyed field.  If a fraction, $\mathcal{F}_{\rm X}$, of the WNSs have
$\sx > S_{\rm X,min}$, as determined from the X-ray flux distribution, {\em
Chandra} should have detected roughly $\langle \tau_{\rm MS} \rangle
\mathcal{R}_{\rm SN} \mathcal{F}_{\rm FE} \mathcal{F}_\Omega \mathcal{F}_{\rm
X}$ such sources.


Here, we define intermediate- and high-mass post-SN binaries by $M_2 =
3$--$8\msun$ and $M_2>8\msun$, respectively.  When we apply the conventional
kick scenario, K1, we obtain $\mathcal{F}_{\rm FE} \sim 0.05$.  The general
importance of intermediate-mass secondaries is indicated by the rather large
mean lifetime of $\langle \tau_{\rm MS} \rangle \sim 50$\,Myr for the post-SN
binaries.  The formation efficiency in kick scenario K2 is $\mathcal{F}_{\rm FE}
\sim 0.2$, and the mean lifetime is reduced to $\langle \tau_{\rm MS} \rangle
\sim 30$\,Myr.  We find that $\mathcal{F}_\Omega \sim 0.01$, so that the total
number of WNSs in the solid angle surveyed by {\em Chandra} is $\sim 250$ and
$\sim 600$ for kick scenarios K1 and K2, respectively.

Examples of the X-ray flux distribution for the WNSs are shown in Fig.~1.  The
two curves in the top panel correspond to our two choices for the wind speed,
both for kick scenario K1.  Kick scenario K2 does not yield an appreciably
different flux distribution, aside from the normalization factor
$\mathcal{F}_{\rm FE}$.  The middle and bottom panels of Fig. 1 show the
contributions of intermediate- and high-mass systems, which change considerably
in the two kick scenarios for reasons given earlier.  The fraction of WNSs with
$S_{\rm X} > S_{\rm X,min}$ depends very much on the prescription
\begin{inlinefigure}
\centerline{\epsfig{file=fhist_031102.epsi,width=\linewidth}}
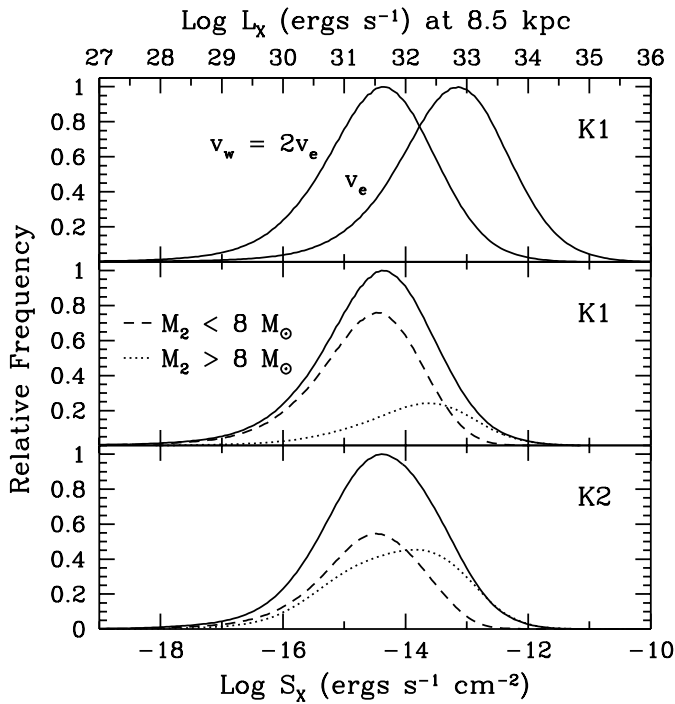
\caption{The X-ray flux distributions for a set of model assumptions.  The top
panel shows the distributions for $v_w = v_e$ ({\em right curve}) and $v_w =
2v_e$ ({\em left curve}), where kick scenario K1 was used.  The middle panel
shows the contributions of intermediate-mass ({\em short-dashed}) and high-mass
({\em dotted}) systems, for $v_e = 2 v_w$ and kick scenario K1.  The bottom
panel is similar to the middle panel, but for kick scenario K2.}
\end{inlinefigure}

\noindent
for the stellar wind speed.  For $v_w = v_e$ and $v_w = 2 v_e$, we find that
$\mathcal{F}_{\rm X} \sim 0.4$ and $\sim 0.05$, respectively, giving {\em
detected} numbers of WNSs between ten and several hundred.

Figure 1 shows that WNSs with intermediate-mass companions may make a dominant
contribution to the flux distribution over a wide range in $\sx$.  This is
largely due to the longer lifetimes and and lower wind speeds for lower
secondary masses.  While only a few of the observed so-called HMXBs
\citep{liu00} have the mid- to late-B spectral types consistent with unevolved
intermediate-mass stars, it is certainly plausible that a much larger number of
the known systems harbor such companions; in most cases the spectral subtype is
not well constrained.  Interestingly, the small fraction of these systems with
$M_2 \la 4\msun$ and $P_{\rm orb} \sim$ 1--$10\day$ will appear as IMXBs
\citep{podsi02} when the secondary fills its Roche lobe and stably transfers
matter to the NS.  For longer periods and larger masses, the mass transfer is
dynamically unstable.  However, as essentially all of the NS binaries with $M_2
\la 8\msun$ form in much the same way, the entire population of WNSs with
intermediate-mass companions may provide unique statistical information
regarding the formation of IMXBs.


\section{DISCUSSION}\label{sec:dis}

{\em Chandra} has detected $\sim 1000$ X-ray sources about the
Galactic center.  The next step is to determine the nature of these sources.
Based on the discussions and results presented in the last two sections, we are
in a position to make suggestions regarding future observational work,
especially as it pertains to WNSs.

A significant fraction --- {\em perhaps the majority} --- of the 3--8-keV point
sources detected by {\em Chandra} may be WNSs located within several kiloparsecs
of the Galactic center.  We propose that an infrared observing campaign be
undertaken to search for stellar counterparts.  Our models indicate that the
majority of WNSs have companions with mid-O to late-B spectral types, with
unreddened $K$-band magnitudes of $\sim 11$--16 at the distance of the Galactic
center, assuming these stars are near the ZAMS.  The intrinsic $JHK$ colors of
hot stars are nearly degenerate, and determination of their spectral subtypes
thus requires a spectral-line classification system \citep[e.g.,][]{hanson96}.
Purely photometric observations are an important first step, however, as they
should allow one to distinguish between stars and background AGN, even with the
effects of interstellar extinction.

We also suggest that it would be worthwhile to extend the {\em Chandra} survey
in both sensitivity and angular coverage.  The X-ray flux distributions shown in
Fig.~1 indicate that a factor of 10 increase in sensitivity may gain a factor of
$\sim 5$--10 in the number of detected WNSs; of course, the number of AGN may
likewise increase by a factor of several tens.  Furthermore, extending the
survey to Galactic latitudes $|b| \ga 2\degr$ may reveal a gradient in the
density of point sources, and thus give the fraction that are Galactic.


\acknowledgements

 We are grateful to F. Baganoff, N. Brandt, D. Chakrabarty, A. Juett, M. Muno,
and Q. D. Wang for useful discussions.  This work was supported in part by NASA
ATP grant NAG5-8368.





\end{document}